\documentclass[conference]{IEEEtran}
\usepackage[none]{hyphenat}

\usepackage{cite}
\usepackage{amsmath,amssymb,amsfonts}
\usepackage{algorithmic}
\usepackage{graphicx}
\usepackage{textcomp}
\usepackage{xcolor}
\def\BibTeX{{\rm B\kern-.05em{\sc i\kern-.025em b}\kern-.08em
    T\kern-.1667em\lower.7ex\hbox{E}\kern-.125emX}}

\def\IEEEkeywordsname{Keywords}

\begin{document}

\title{Deep Learning for Lung Disease Classification Using Transfer Learning and a Customized CNN Architecture with Attention\\

}

\author{
    \IEEEauthorblockN{Xiaoyi Liu}
    \IEEEauthorblockA{\textit{Ira A. Fulton Schools of Engineering} \\
        \textit{Arizona State University}\\
        Tempe, USA \\
        xliu472@asu.edu}
    \and
    \IEEEauthorblockN{Zhou Yu}
    \IEEEauthorblockA{\textit{Department of Mathematics, Statistics,} \\
        \textit{and Computer Science} \\
        \textit{University of Illinois at Chicago}\\
        Chicago, USA \\
        zyu33@uic.edu}
    \and
    \IEEEauthorblockN{Lianghao Tan}
    \IEEEauthorblockA{\textit{W. P. Carey School of Business} \\
        \textit{Arizona State University}\\
        Tempe, USA \\
        ltan22@asu.edu}
}

\maketitle

\begin{abstract}
Many people die from lung-related diseases every year. X-ray is an effective way to test if one is diagnosed with a lung-related disease or not. This study concentrates on categorizing three distinct types of lung X-rays: those depicting healthy lungs, those showing lung opacities, and those indicative of viral pneumonia. Accurately diagnosing the disease at an early phase is critical. In this paper, five different pre-trained models will be tested on the Lung X-ray Image Dataset. SqueezeNet, VGG11, ResNet18, DenseNet, and MobileNetV2 achieved accuracies of 0.64, 0.85, 0.87, 0.88, and 0.885, respectively. MobileNetV2, as the best-performing pre-trained model, will then be further analyzed as the base model. Eventually, our own model, MobileNet-Lung based on MobileNetV2, with fine-tuning and an additional layer of attention within feature layers, was invented to tackle the lung disease classification task and achieved an accuracy of 0.933. This result is significantly improved compared with all five pre-trained models.
\end{abstract}


\section{Introduction}
Lung disease is a major contributor to global morbidity, encompassing chronic conditions such as chronic obstructive pulmonary disease (COPD), asthma, and lung cancer, alongside acute infections like viral pneumonia and tuberculosis \cite{jadhav2021introduction}. Early and accurate diagnosis of these diseases is critical, as it can dramatically improve patient outcomes. X-rays are one of the major tools used by doctors to diagnose lung disease. However, the massive volume of X-ray images and the complexity of each individual case can make it challenging for doctors to achieve the best performance.

Deep learning (DL) has made significant achievements in various industries, including healthcare and medical image classification. This paper will use five different pre-trained models to test the best way to predict possible lung diseases using the publicly available Lung X-ray Image Dataset. Among all five models, MobileNetV2 achieved the best accuracy of 0.885 compared with SqueezeNet, VGG11, ResNet18, and DenseNet, which had accuracies of 0.64, 0.85, 0.87, and 0.88, respectively.

MobileNetV2 was then used as the base model. Fine-tuning was tested on MobileNetV2 with a specific dataset for lung disease classification tasks, resulting in a significant improvement in accuracy to 0.925. Finally, a new model was introduced based on the fine-tuned MobileNetV2 model. An additional layer using an attention mechanism within the feature layer further improved the model's performance, achieving an accuracy of 0.933.

\section{Background}
\subsection{Machine Learning}

Machine learning is essential to many modern advancements and applications. Many applications, including robot collaboration \cite{liu2024enhanced}, credit card fraud detection \cite{yu2024credit}, heterogeneous data environments \cite{wu2024application}, autonomous robot navigation \cite{wang2024research}, vehical classification \cite{202407.0981}, threat detection \cite{Weng202404}, and payment security systems \cite{zheng2024advanced}, are enhanced by machine learning.

\subsection{Deep Learning}

Deep learning technologies, as a subfield of machine learning, are widely applied in various fields today, including natural language processing (NLP), computer vision (CV), and more \cite{lecun2015deep}. For CV, applications like real-time pill identification \cite{dang2024real}, snore sound analysis \cite{article}, and deepfake detection \cite{lai2024gm, lai2024selective} are good examples. This paper will concentrate on applying deep learning techniques for medical image classification related to lung diseases.

\subsection{CNN}
Convolutional neural networks (CNNs) play a vital role in CV. Many CNN models have been developed for various CV tasks. In this paper, five pre-trained models—SqueezeNet, DenseNet, MobileNetV2, ResNet18, and VGG11—will be analyzed for lung disease classification tasks.

\subsection{Lung Opacity and Viral Pneumonia}

Lung opacities on X-rays can indicate different diseases such as pulmonary edema, lung cancer, and bacterial pneumonia. These opacities appear as hazy areas that, while denser than normal lung tissue, do not obscure the underlying structures but disrupt the normal lung architecture \cite{turk2024detection}. Viral pneumonia is triggered by viruses like influenza and respiratory syncytial virus (RSV). Viral pneumonia X-rays show diffuse, bilateral involvement, often with a "ground-glass" appearance, indicating partial filling of airspaces with exudate or other material \cite{stefanidis2021radiological}. Being able to identify the differences between lung opacities, normal lungs, and viral pneumonia from X-ray images can help doctors determine the primary cause of the lung disease, leading to better and earlier treatments for patients, which can be critical. Figure 1 shows X-ray images depicting normal lungs, lung opacity, and viral pneumonia.

\begin{figure}[ht]
    \centering
    \includegraphics[width=1\linewidth]{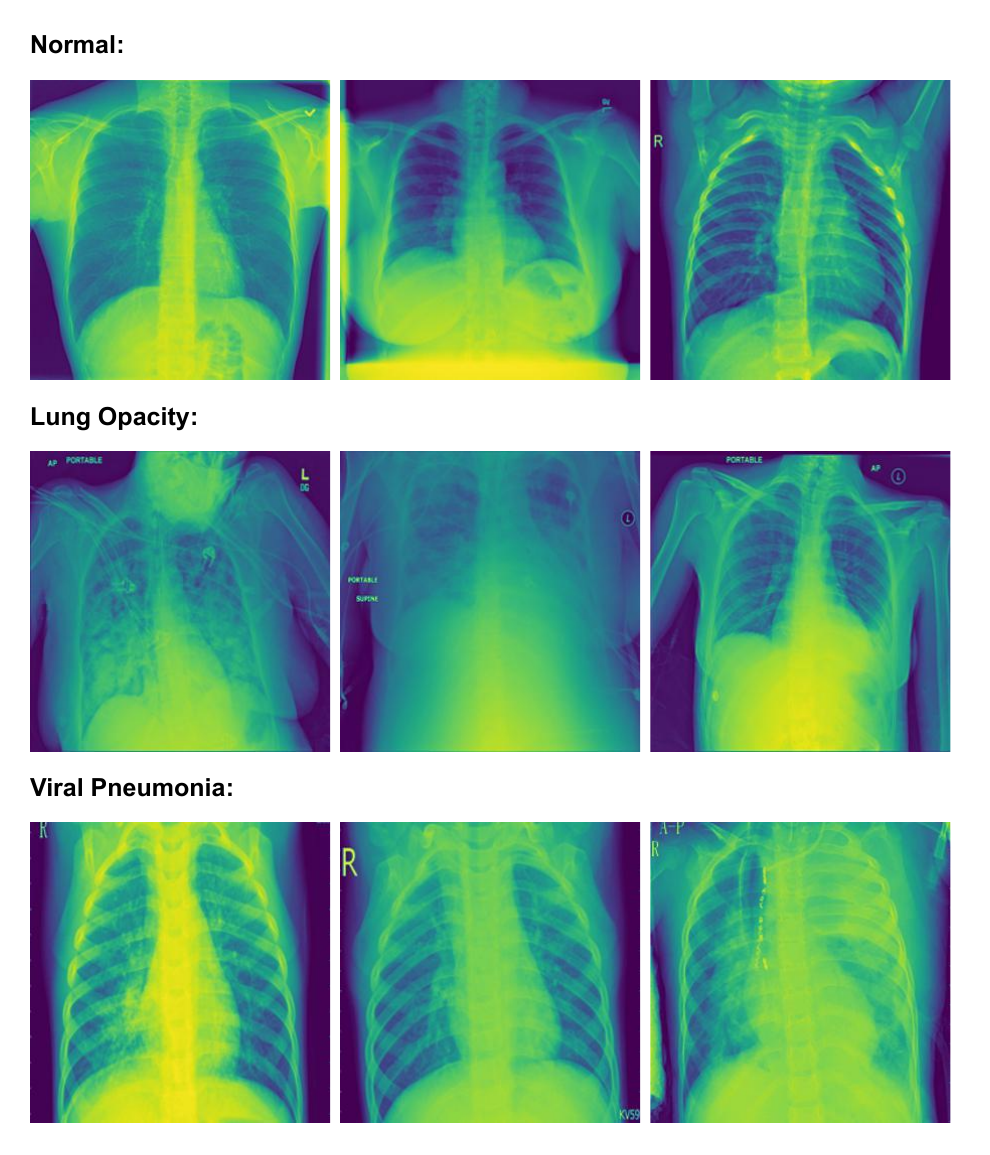}
    \caption{Lung Disease Types}
    \label{fig: Lung Disease Types}
\end{figure}

\subsection{Pre-trained Models}

\subsubsection{SqueezeNet}
SqueezeNet is a lightweight CNN model designed to achieve AlexNet-level accuracy using significantly fewer parameters. The use of Fire modules dramatically reduces the parameter count. It is ideal for mobile and other computationally limited platforms due to its fast reaction time \cite{iandola2016squeezenet}.

\subsubsection{VGG11}

VGG11 is part of the VGG family. It consists of 11 layers, including convolutional and dense layers. 3x3 convolution filters are stacked to increase the depth of the model \cite{simonyan2014very}. VGG11 has been widely used in many applications and demonstrates the importance of using deeper models for better performance.

\subsubsection{ResNet-18}

ResNet introduced the concept of residual learning. It allows the network to skip certain layers, enabling the model to go deeper and create more complex architectures. This approach remedies the vanishing gradient problem, allowing for more effective training of deeper networks \cite{he2016deep}. ResNet ensures information propagates through the network efficiently.

\subsubsection{DenseNet}

DenseNet is another pre-trained CNN model that will be tested in this study. In DenseNet, each layer receives inputs from all previous layers, which enhances feature propagation and reduces the number of parameters \cite{huang2017densely}. It is inspired by, and one step further than, ResNet.

\subsubsection{MobileNetV2}
MobileNetV2 is primarily designed for mobile devices and other platforms with limited computational resources. As there is a large amount of medical image classification work in hospitals, a quick analysis is critical. Inverted residuals and linear bottlenecks are introduced in MobileNetV2, making it ideal for real-time image classification tasks with high performance and high efficiency \cite{sandler2018mobilenetv2}.

\begin{table*}[ht!]
\centering

\caption{Comparison of Models on Lung Disease Classification}
\label{table:results}
\begin{tabular}{lcccccc}
\hline
\textbf{Model} & \textbf{Ave. Loss} & \textbf{Accuracy} & \textbf{Precision} & \textbf{Recall} & \textbf{F1-Score} \\
\hline
SqueezeNet & 0.8840 & 0.64 & 0.7189 & 0.64 & 0.5565 \\ 

VGG11 & 0.5268 & 0.8507 & 0.8491 & 0.8507 & 0.8498 \\ 

ResNet18 & 0.3191 & 0.8693 & 0.8694 & 0.8693 & 0.8685 \\ 

DenseNet & 0.3470 & 0.88 & 0.8834 & 0.88 & 0.8810 \\ 

MobileNetV2 & 0.3033 & 0.8853 & 0.8864 & 0.8853 & 0.8857 \\ 

MobileNet-FT & 0.2557 & 0.9253 & 0.9281 & 0.9253 & 0.9248 \\ 

\textbf{MobileNet-Lung (Ours)} & \textbf{0.2305} & \textbf{0.9333} & \textbf{0.9337} & \textbf{0.9333} & \textbf{0.9332} \\

\hline
\end{tabular}

\end{table*}

\section{Methodology}

\subsection{Dataset}

The Lung X-Ray Image Dataset, a publicly available dataset, is employed for this research \cite{talukder2023lung}. It consists of 3475 X-ray images. Among these, 1250 images are normal lung X-ray images, indicating a healthy lung condition. There are 1125 images of lung opacity X-ray images, indicating different degrees of lung abnormalities. Additionally, there are 1100 images of viral pneumonia X-ray images, demonstrating lung infections related to viral pneumonia. The dataset is divided into 3 subsets: 80\% for the training dataset, 10\% for the validation dataset, and 10\% for the testing dataset.

\subsection{Image Augmentation and Preprocessing}

Image augmentation and preprocessing are critical processes to help the model achieve better results. The training transformations include random horizontal and vertical flips with a probability of 50\%, random rotations up to 10 degrees, resizing to 224 x 224 image size, and converting to a tensor that ranges from 0 to 1. Normalization is performed using specific mean values and standard deviations tailored to the dataset to ensure consistency in image preprocessing.

\subsection{Learning Rate and Epoch}

The learning rate starts at 0.01 with a momentum of 0.9. The Stochastic Gradient Descent optimizer is used in this study. To achieve better results and convergence, the learning rate is multiplied by 0.1 every 10 epochs. Training will be halted if validation accuracy does not improve for 10 consecutive epochs.

\subsection{Pre-trained Models}
Five different pre-trained models are tested in this study: SqueezeNet, ResNet18, VGG11, DenseNet121, and MobileNetV2. They achieve accuracies of 0.64, 0.85, 0.86, 0.88, and 0.885, respectively, on the testing dataset. When comparing the size of the models, they have 1.25 million, 11.7 million, 132.9 million, 7.98 million, and 3.5 million parameters, respectively. Since MobileNetV2 achieves the best result and has the second least number of parameters, further studies will be based on the MobileNetV2 model. More information on average loss, precision, F1-score, and recall are presented in Table I.

\begin{figure}[!ht]
    \centering
    \includegraphics[width=1\linewidth]{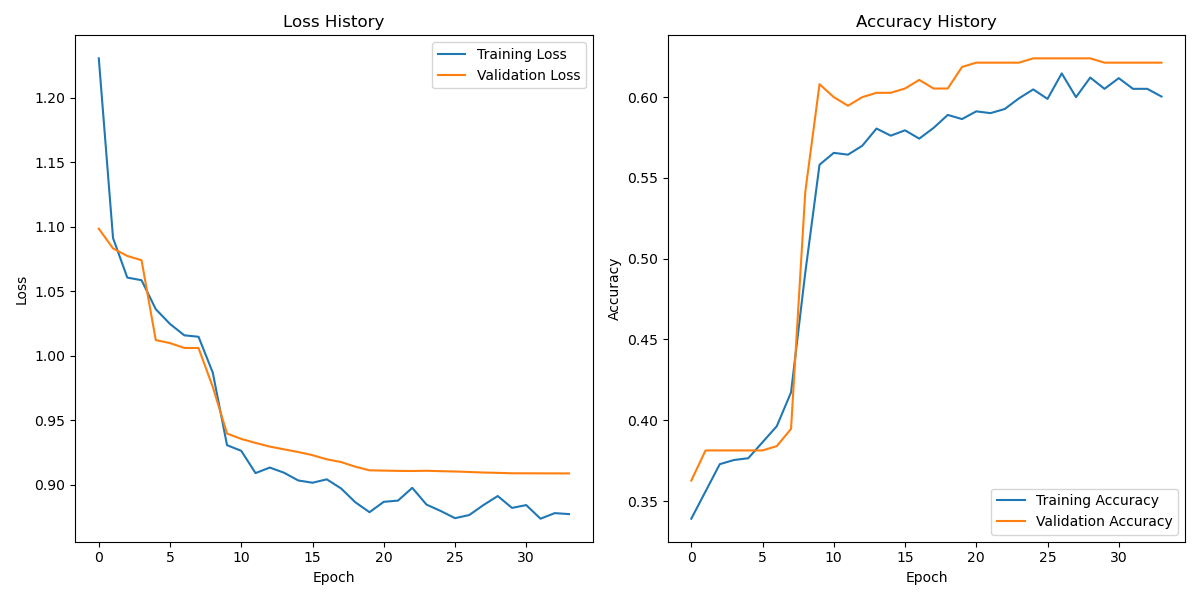}
    \caption{SqueezeNet Loss VS Accuracy}
    \label{fig:SqueezeNet Loss VS Accuracy}
\end{figure}

\begin{figure}[!ht]
    \centering
    \includegraphics[width=1\linewidth]{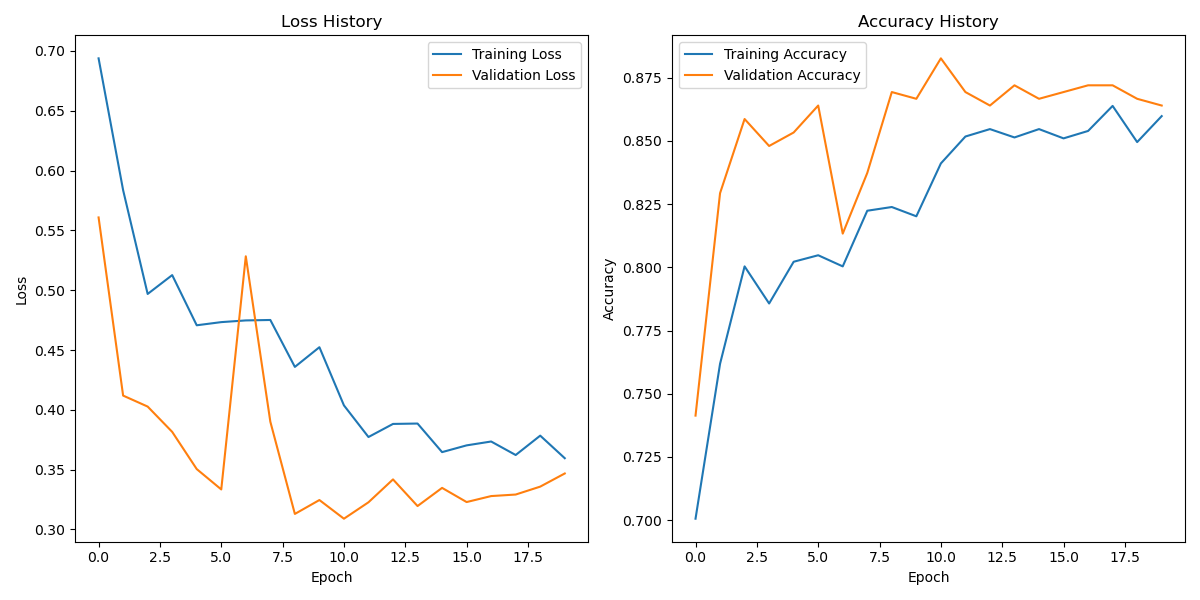}
    \caption{ResNet18 Loss VS Accuracy}
    \label{fig: ResNet18 Loss VS Accuracy}
\end{figure}

\begin{figure}[!ht]
    \centering
    \includegraphics[width=1\linewidth]{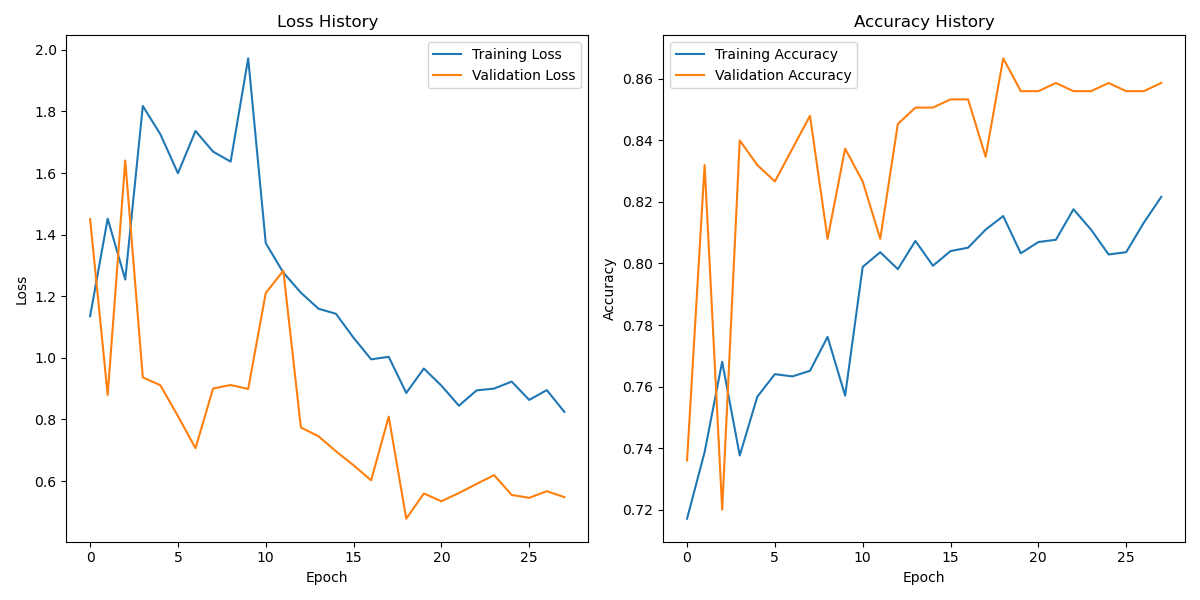}
    \caption{VGG11 Loss VS Accuracy}
    \label{fig: VGG11 Loss VS Accuracy}
\end{figure}

\begin{figure}[!ht]
    \centering
    \includegraphics[width=1\linewidth]{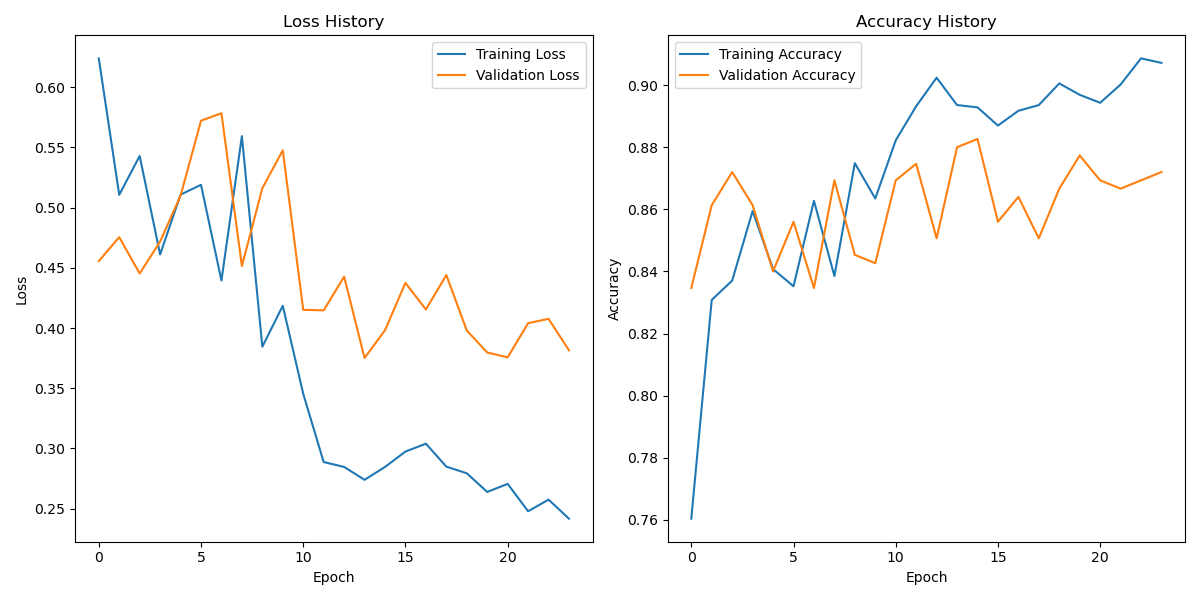}
    \caption{DenseNet Loss VS Accuracy}
    \label{fig: DenseNet Loss VS Accuracy}
\end{figure}

\begin{figure}[!ht]
    \centering
    \includegraphics[width=1\linewidth]{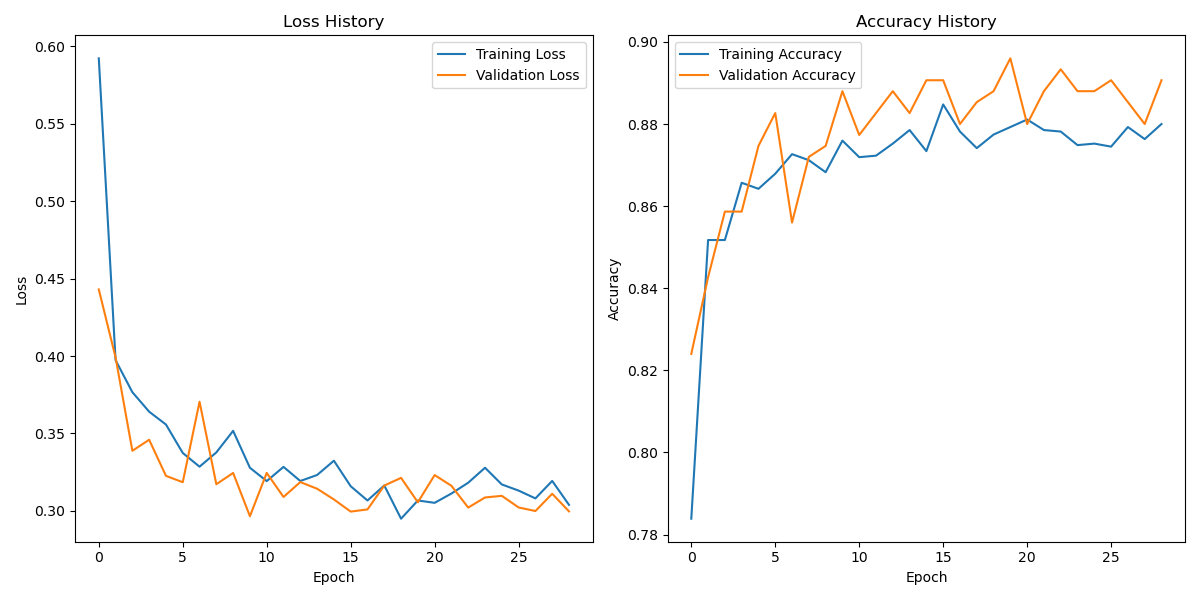}
    \caption{MobileNetV2 Loss VS Accuracy}
    \label{fig: MobileNet_V2 Loss VS Accuracy}
\end{figure}

\subsection{Attention}

The attention mechanism is crucial in deep learning. It mimics the biological systems which will focus on the most important part of the information rather than the whole information \cite{niu2021review}. The attention mechanism is widely applied in various areas, including CV, speech recognition, and other fields. This study focuses on the application of attention mechanisms in the field of CV.

More specifically, the Squeeze-and-Excitation (SE) block will be applied in this study. The SE block introduces a novel way to improve CNNs by using two main operations: the squeeze operation and the excitation operation \cite{hu2018squeeze}. 

The squeeze operation aims to exploit the relationships between channels in CNNs. Each learned filter in the network processes data within a local receptive field, avoiding the need to capture broader contextual information \cite{hu2018squeeze}. This is done using global average pooling, which computes statistics for each channel \cite{hu2018squeeze}. For an output feature map \( U \) with spatial dimensions \( H \times W \) and \( C \) channels, the \( c \)-th element of the channel descriptor \( z \) is calculated as:

\begin{equation}
z_c = F_{sq}(u_c) = \frac{1}{H \times W} \sum_{i=1}^{H} \sum_{j=1}^{W} u_c(i, j)
\end{equation}

Following the squeeze operation, the SE block utilizes an "excitation" operation to capture dependencies across channels \cite{hu2018squeeze}. This step models nonlinear interactions between channels and highlights multiple channels using a gating mechanism activated by a sigmoid function:

\begin{equation}
s = F_{ex}(z, W) = \sigma(g(z, W)) = \sigma(W_2 \delta(W_1 z)),
\end{equation}

In this context, \(\delta\) represents the ReLU function, while \(W_1\) belongs to \(\mathbb{R}^{\frac{C}{r} \times C}\) and \(W_2\) belongs to \(\mathbb{R}^{C \times \frac{C}{r}}\).

The final output of the SE block is achieved by rescaling \(U\) using the activations \(s\):

\begin{equation}
x_{ec} = F_{scale}(u_c, s_c) = s_c \cdot u_c,
\end{equation}

In this equation, it can be observed that \(X_e = [x_{e1}, x_{e2}, \ldots, x_{eC}]\) and \(F_{scale}(u_c, s_c)\) involves channel-wise multiplication between the feature map \(u_c \in \mathbb{R}^{H \times W}\) and the scalar \(s_c\) \cite{hu2018squeeze}.

\subsection{Proposed Deep Learning Models}
Because MobileNetV2 achieved the highest accuracy and F1-score with a relatively small model parameter size among all five pre-trained models, further analysis and testing were based on the MobileNetV2 model. After several rounds of testing, it was determined that the best way to fine-tune the pre-trained MobileNetV2 architecture is to unfreeze all layers. This newly fine-tuned model, named MobileNet-FT, achieved an accuracy of 0.925 compared to MobileNetV2's accuracy of 0.885. This represents a significant improvement from MobileNetV2's accuracy. Although pre-trained models are trained on much larger datasets like ImageNet, it is preferable to fine-tune them based on specific datasets, especially for tasks, such as medical imaging. Utilizing the initial weights and biases of a pre-trained model accelerates the training process.

Afterward, a model called MobileNet-Lung was developed. The process began by modifying the MobileNet-FT model, which had been fine-tuned. An SE block was incorporated after the first convolutional layer, while the rest of the feature layers remained unchanged. By doing so, the new model leverages the benefits of SE blocks while maintaining the efficiency and high accuracy of MobileNet-FT. Additionally, the classifier was adjusted to match the number of output classes for the specific task of lung classification. After all the updates, the new model achieved an increased accuracy and F1-score on the testing dataset, reaching 0.933 and 0.9332, respectively, compared to MobileNet-FT's 0.925 and 0.924.

\subsection{Performance Measurements}
For tasks such as classification, accuracy and F1-score are good metrics. In addition to these, precision and recall are also important measurements.

Other than those metrics, the average loss is used as a measurement. In this study, cross-entropy loss was employed. Cross-entropy loss is particularly useful for classification tasks. Here, \(N\) is the number of classes, \(y_i\) is the true label, and \(\hat{y}_i\) is the predicted probability for class \(i\). The cross-entropy function is shown below:

\begin{equation}
L = -\sum_{i=1}^{N} y_i \log(\hat{y}_i)
\end{equation}

These metrics and loss functions offer a comprehensive understanding of the model's performance. More comparisons and testing results of different pre-trained models and customized models can be found in Table I.

\begin{figure}[!ht]
    \centering
    \includegraphics[width=1\linewidth]{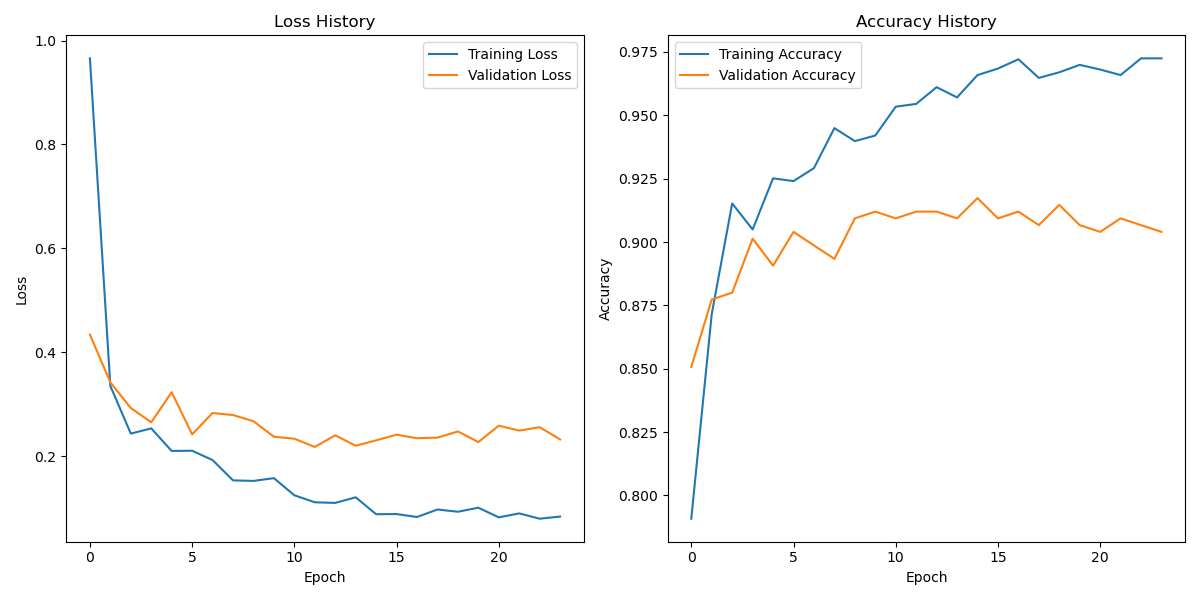}
    \caption{MobileNet-FT Loss VS Accuracy}
    \label{fig: MobileNet-FT Loss VS Accuracy}
\end{figure}

\begin{figure}[!ht]
    \centering
    \includegraphics[width=1\linewidth]{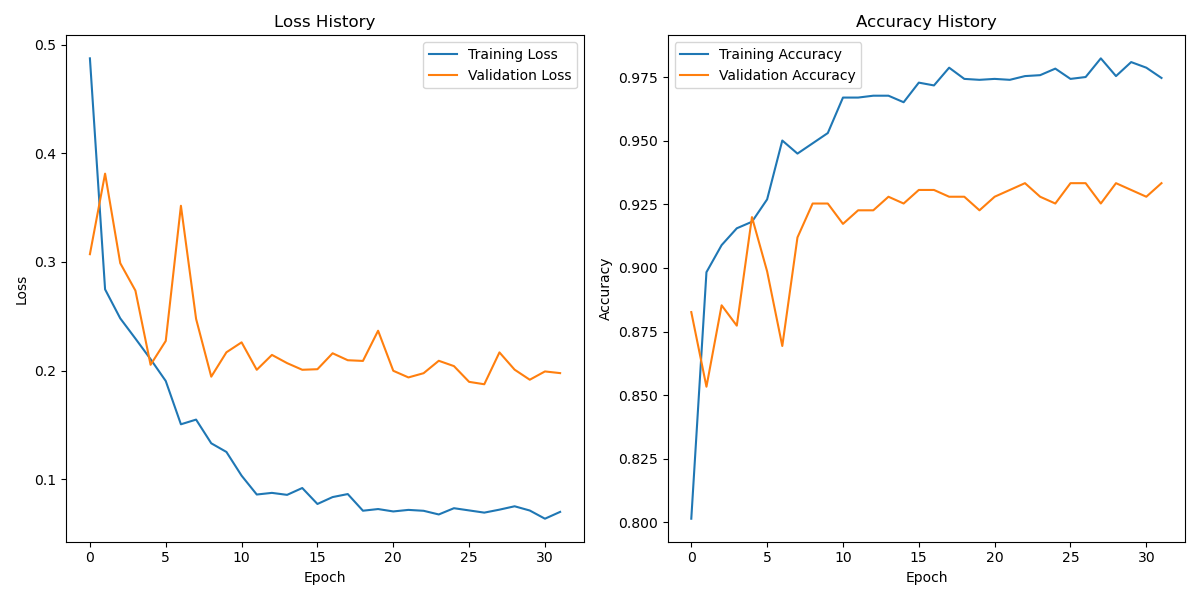}
    \caption{MobileNet-Lung Loss VS Accuracy}
    \label{fig: MobileNet-Lung Loss VS Accuracy}
\end{figure}

\begin{figure}[!ht]
    \centering
    \includegraphics[width=1\linewidth]{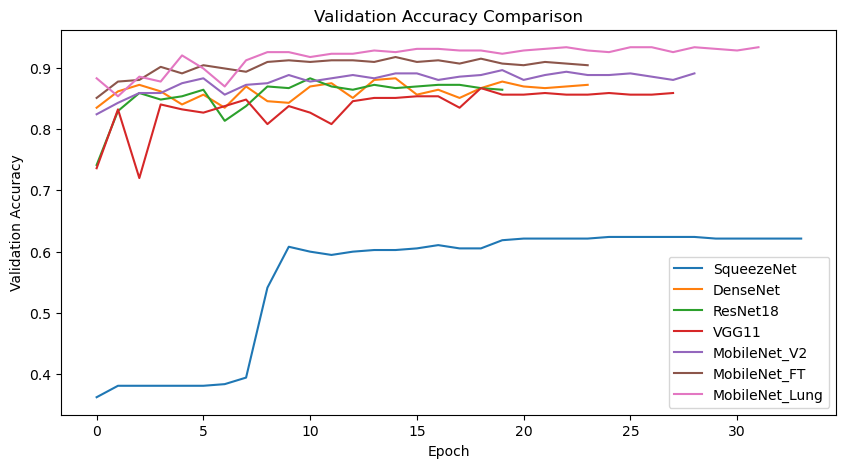}
    \caption{Validation Accuracy Comparison}
    \label{Validation Accuracy Comparison}
\end{figure}

\section{Evaluation and Discussion of Results}

First, let's examine the performance of the five pre-trained models. Figures 2, 3, 4, 5, and 6 illustrate the average loss and accuracy over epochs for each model. Among these models, MobileNetV2 delivered the best results, achieving an accuracy of 0.885 and very smooth convergence curves. Conversely, the worst-performing model, SqueezeNet, only achieved an accuracy of 0.64.

Next, the customized models were evaluated. MobileNet-FT, as shown in Figure 7, is fine-tuned based on MobileNetV2. This model achieved higher accuracy and lower average loss, with an accuracy of 0.925 and an average loss of 0.2557, compared to the best-performing pre-trained model MobileNetV2, which had an accuracy of 0.885 and an average loss of 0.303. These results underscore the necessity of fine-tuning pre-trained models for specific tasks, such as medical image classification.

For our novel model, MobileNet-Lung, as shown in Figure 8, an SE block was incorporated into the MobileNet-FT model after the first convolutional layer. This modification resulted in the highest accuracy of 0.933, compared to MobileNet-FT's 0.925, and the lowest average loss of 0.23, compared to MobileNet-FT's 0.2557. This improvement suggests that the additional attention mechanism enables the model to concentrate more effectively on crucial features, thereby enhancing performance.

Figure 9 clearly shows that MobileNet-Lung achieved the highest accuracy on the validation dataset. MobileNet-FT is a close second. However, MobileNet-Lung requires approximately 33 epochs to train, whereas MobileNet-FT requires roughly 23 epochs. Therefore, when dealing with large datasets or limited computational resources, MobileNet-FT may be a more suitable option.

\section{Conclusion}

For the lung disease classification task, five different pre-trained models were tested, achieving accuracies of 0.64, 0.85, 0.869, 0.88, and 0.885, respectively. MobileNetV2, as the best-performing model, was then used as the base model for further analysis. After fine-tuning, the model, referred to as MobileNet-FT, achieved an accuracy of 0.925. Finally, a new model called MobileNet-Lung was introduced for the specific task of lung disease classification, achieving an accuracy of 0.933. The attention mechanism of the SE block and fine-tuning the pre-trained model dramatically increased the performance of medical image classification tasks. When computational resources are limited, MobileNet-FT is the more suitable model for better efficiency with a minor accuracy drop.

\bibliography{reference.bib}
\bibliographystyle{IEEEtran}

\vspace{12pt}
\end{document}